\documentclass[	
						superscriptaddress,%
						twocolumn,%
						a4paper,
						showkeys,
					]{revtex4}

\usepackage[frenchb,english]{babel}
\usepackage[T1]{fontenc}
\usepackage{upgreek}
\usepackage{textcomp} 				
\usepackage[dvips]{graphicx}		
\usepackage{color}					
\usepackage{amssymb,amsmath}		
\usepackage{array}					
\usepackage{ulem}

\renewcommand{\emph}[1]{\textit{#1}}
\definecolor{vert}{rgb}{0.5,0.758,0.5}
\definecolor{bleufonce}{rgb}{0,0,0.516}
\definecolor{orange}{rgb}{1,0.516,0}

\begin{document}

\title{Order-Parameter Symmetries of Domain Walls in Ferroelectrics and Ferroelastics}
\date{\today}
\author{Pierre Tol\'edano}
\affiliation{Laboratoire des Syst\`emes Complexes, Universit\'e de Picardie, 80000 Amiens, France}
\author{Mael Guennou}
\email{guennou@lippmann.lu}
\affiliation{Département Science et Analyse des Matériaux, CRP Gabriel Lippmann, 41 rue du Brill, L-4422 Luxembourg}
\author{Jens Kreisel}
\affiliation{Département Science et Analyse des Matériaux, CRP Gabriel Lippmann, 41 rue du Brill, L-4422 Luxembourg}
\affiliation{Physics and Materials Science Research Unit, University of Luxembourg, 41 Rue du Brill, L-4422 Belvaux, Luxembourg}

\begin{abstract}
The symmetry of boundaries between ferroelectric, ferroelastic, and antiphase domains is a key element for a theoretical understanding of their properties. Here, we derive this symmetry from their organic relation to the symmetry of the primary transition order parameters. The domain wall symmetries are shown to coincide with directions of the order-parameter $n$-dimensional vector space, corresponding to sum of the vectors associated with adjacent domain states. This property is illustrated by the determination of the maximal symmetries of domain walls in BaTiO$_3$, LaAlO$_3$, SrTiO$_3$ and Gd$_2$(MoO$_4$)$_3$. Besides, the domain pattern in YMnO$_3$ is interpreted as resulting from an annihilation-creation process,  the annihilation of the antiphase domain walls creating six ferroelectric domain walls merging at a single point.
\end{abstract}


\maketitle

\section{Introduction}

Multiferroic materials, i.e., materials possessing simultaneously several ferroic orders such as ferromagnetism, ferroelectricity, and/or ferroelasticity, currently attract a great deal of interest because of their intriguing coupling phenomena. One of the interesting and intrinsic properties of ferroic materials is the presence of domains separated by domain walls (DWs, also called domain boundaries) \cite{Wadhawan2000}. DWs can be seen as spatially extended transition regions mediating the change in the ferroic order parameters from one domain to another with resulting gradient effects. For a long time, however, ferroelectric and ferroelastic DWs have been considered experimentally as interfaces with negligible thickness. It is only within the last few years, thanks to now available atomic-resolution studies such as high-resolution transmission electron microscopy, various atomic force microscopies (c-AFM, PFM), etc. that their real complexity has been revealed. This has inspired a new paradigm of ferroic devices where the domain wall, rather than the domain, is the active element. It has been argued that the exploitation of the small size of domain walls (of the order of several nanometers) and their different functional properties present a high potential towards domain wall nanoelectronics \cite{Bea2009a,Salje2010,Catalan2012}. 

Experimentally, pioneering work on WO$_3$ \cite{Aird1998} has shown that ferroelastic twin walls exhibit electric conductivity or even superconductivity. Conducting domain walls were observed in several materials, e.g., BiFeO$_3$ \cite{Seidel2009} or BaTiO$_3$ \cite{Sluka2013}, which exhibit electronic conductivity up to 10$^9$ higher than the insulating domains. Interestingly, in other ferroelectric materials such as in LiNbO$_3$ single crystals, the domain-wall conductivity is induced only by super-bandgap UV-light illumination \cite{Schroeder2012}. Ferroelastics such as CaTiO$_3$ \cite{VanAert2012} or SrTiO$_3$ \cite{Salje2013} present the striking example of domain walls with polar properties while the bulk material is not ferroelectric. The full understanding and engineering of such domain wall properties remain to be established. In particular, it is not yet clear if all reported results reflect the intrinsic properties of the domain walls, or if they are related to the complex minimization of the different energy ingredients, such as surface energy, residual stresses, shape anisotropy, structural defects, impurities or stoichiometry issues contributing to their stabilization. As a consequence of this, theoretical tools are needed to distinguish intrinsic from extrinsic effects and properties, symmetry being thereby an essential ingredient. The symmetry properties of DW have already been subjected to theoretical investigations. Classically, the symmetry of the domain wall can be deduced by combining the common symmetry elements of the adjacent domains plus additional symmetries transforming one domain into another \cite{Janovec2006}. Notably, following this approach, the symmetries of strain-compatible walls in ferroelectrics and ferroelastics has been determined and tabulated \cite{Janovec2006,Erhart2004}. 

Here, we propose an alternative approach which allows deriving the symmetry of the domain walls from their organic relationship with the primary transition order parameter, giving rise to the domain pattern. We show that although in most cases the symmetry of domain walls as obtained by geometrical considerations \cite{Janovec2006} can be straightforwardly deduced from the symmetry of the corresponding order parameter, in a number of specific situations, considering or ignoring the order parameter symmetry leads to different predictions.

\section{General approach}

Our proposed approach for determining domain-wall symmetries is based on the following property: \emph{The symmetry of the domain wall between two adjacent domains associated with the vectors $\vec V_1$ and $\vec V_2$ in the $n$-dimensional order-parameter vector space is an isotropy subgroup of the symmetry group of the domain state corresponding to $\vec V_1+\vec V_2$.} This is inferred from the fact that the symmetry group of a domain wall leaves by definition the domain pair invariant: the common symmetry operations of adjacent domains represented by $\vec V_1$ and $\vec V_2$ as well as the operations exchanging the two vectors also leave their sum invariant. Therefore, the symmetry group $G_0$ of $\vec V_1+\vec V_2$ contains the isotropy subgroup $G$ corresponding to the symmetry of the domain wall. $G$ may coincide with $G_0$ or be a proper isotropy subgroup of $G_0$. 

The symmetry $G$ of the domain wall as determined from our approach is described by a point group. It is a maximal symmetry that is independent of a particular choice of domain wall orientation, which in practice may result from elastic compatibility, from the history and preparation of the sample or the minimization of bound charges. A particular choice of orientation requires us to take into account the planar character of the wall in space and describe the DW symmetry by a layer group \cite{Janovec2006}. A symmetry lowering generally occurs in the process, which is linked to the geometric arrangement of domains in space and the DW orientation but does not follow from the phase transition mechanism itself. Besides, when assuming a finite wall thickness, the DW symmetry may be further reduced if a phase transition occurs \emph{within} the DW, with the emergence of another stable state associated with the same order parameter inducing the transition in the bulk \cite{Bulbich1988}. Such structural phase transitions have not yet been observed in the domain walls of ferroelectric or ferroelastic materials but have been predicted theoretically \cite{Lajzerowicz1979,Sonin1989,Lee2009,Yudin2012,Stepkova2012} and reported experimentally in domain walls separating magnetic domains \cite{Zalesskii1975,Bogdanov1984,Galkina1986}. Such considerations, which are inherent to the metastable nature of domains and domain walls due to their positive energy, concern a marginal number of experimental situations, and do not limit the interest of our general approach. 

In the following, we shall apply this method to model systems representative of the different situations occurring in ferroelectric and ferroelastics: (i) the proper ferroelectric BaTiO$_3$, (ii) the ferroelastics SrTiO$_3$ and LaAlO$_3$, and (iii) the improper ferroelectrics Gd$_2$(MoO$_4$)$_3$ and YMnO$_3$.

\section{Domain-wall symmetry in ferroelectric B\MakeLowercase{a}T\MakeLowercase{i}O$_3$}

\begin{table}
\begin{tabular}{l >{$}l<{$} >{$}l<{$} c l >{$}l<{$} >{$}l<{$}}
\hline\hline
\multicolumn{3}{c}{BaTiO$_3$} &$\ $& \multicolumn{3}{c}{SrTiO$_3$,LaAlO$_3$}\\
\cline{1-3}\cline{5-7}
0 	& (0,0,0) 			& Pm\overline 3m 	&& 0 	& (0,0,0) 									& Pm\overline 3m 	\\
I 	& (P,0,0) 			& P4mm 						&& I 	& (\eta,0,0) 								& I4/mcm 					\\
II 	& (P,P,0) 			& Amm2 						&& II 	& (\eta,\eta,0) 					& Imma 						\\
III & (P,P,P) 			& R3m 						&& III & (\eta,\eta,\eta) 				& R\overline 3c 	\\
IV 	& (P_x,P_x,P_z) & Pm 							&& IV 	& (\eta_1,\eta_2,0) 			& C2/m 						\\
V 	& (P_x,P_y,0) 	& Cm 							&& V 	& (\eta_1,\eta_1,\eta_3) 		& C2/c 						\\
VI 	& (P_z,P_z,P_x) & Cm 							&& VI 	& (\eta_3,\eta_3,\eta_1) 	& C2/m 						\\
VII & (P_x,P_y,P_z) & P1 							&& VII & (\eta_1,\eta_2,\eta_3) 	& P\overline 1 		\\
\hline\hline
\end{tabular}
\caption{Equilibrium states induced by the three-component polarization order parameter associated with the ferroelectric transitions in BaTiO$_3$ and the improper ferroelastic transitions in LaAlO$_3$ and SrTiO$_3$: labeling of the state, equilibrium values of the order-parameter components for one domain state, space-group of the domain state.}
\label{tab1}
\end{table}

\begin{figure}
\begin{center}
\includegraphics[width=0.48\textwidth]{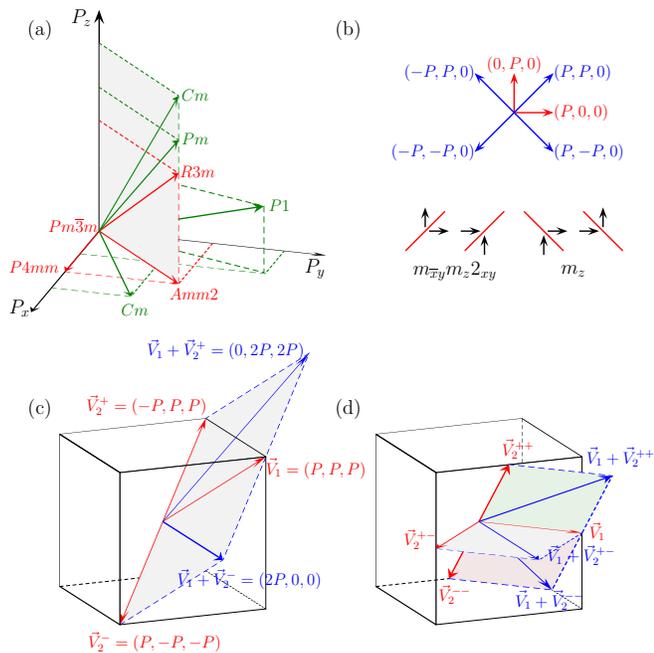}
\caption{(a) Directions representing one domain of the equilibrium states induced by the three-dimensional order-parameter of BaTiO$_3$. Red arrows correspond to the observed states. (b)-(d) Vectors associated with domain walls (blue arrows) between adjacent domains (red arrows) in the tetragonal (b), rhombohedral (c) and orthorhombic (d) phases of BaTiO$_3$. The detailed meaning of the figures is in the text. In Fig. (b) black arrows show the orientations of the polarization on both sides of a domain walls.}
\label{fig1}
\end{center}
\end{figure}

As a first illustrative example of our approach, we consider the ferroelectric domains of the three ferroelectric-ferroelastic phases of barium titanate BaTiO$_3$ having the $4mm$, $3m$, and $mm2$ point groups \cite{LinesAndGlass}. Table \ref{tab1} lists the set of stable states associated with the primary polarization order parameter and the corresponding symmetries of each domain state \cite{Stokes1988,Toledano1996}. Figure \ref{fig1} (a) shows the vectors associated with one domain of each state in the three-dimensional order-parameter space $E_3$. 

In the tetragonal phase the sum of the vectors of $E_3$ associated with a pair of ferroelectric domains at 90$^\circ$ coincides with the symmetry of the orthorhombic domain states. Choosing, e.g., $\vec V_1^{\pm}=(\pm P,0,0)$ and $\vec V_2^{\pm}=(0,\pm P,0)$, we have four possible combinations and two equivalent point-group symmetries, $m_{\overline xy}m_z2_{xy}$ and $m_{xy}m_z2_{\overline xy}$, corresponding to the symmetry of $\vec V_1^{\pm}+\vec V_2^{\pm}=(\pm P,\pm P,0)$  (Table \ref{tab1}). Depending on the orientation of the domain wall, each sum can yield the four configurations (head-to-head, tail-to-tail, tail-to-head and head-to-tail) observed experimentally \cite{Merz1954}, as illustrated in Fig. \ref{fig1} (b) for the $(P,P,0)$ case. For head-to-tail configurations, the symmetry reduces from $m_{\overline xy}m_z2_{xy}$ to $m_z$.

The domain wall symmetries separating the 71$^\circ$ and 109$^\circ$ rhombohedral domains of BaTiO$_3$ are obtained in the same way: putting $\vec V_1=(P,P,P)$ and $\vec V_2^+=(-P,P,P)$ the sum $\vec V_1+\vec V_2^+=(0,2P,2P)$ corresponds to the orthorhombic domain state of symmetry $m_xm_{\overline yz}2_{yz}$ providing the 71$^\circ$-domain wall symmetry (Fig. \ref{fig1} (c)). With $\vec V_2^-=(P,-P,-P)$, the sum $\vec V_1+\vec V_2^-=(2P,0,0)$ coincides with the direction of the tetragonal domain state of symmetry $4_xmm$, the 109$^\circ$-domain wall symmetry corresponding to $m_{yz}m_{\overline yz}2_{x}$ (Fig. \ref{fig1} (c)), which is a maximal isotropy subgroup of $4_xmm$. 

Neighbouring orthorhombic domains present a variety of domain wall symmetries. For the domain states $\vec V_1=(P,P,0)$ and $\vec V_2^{+-} = (P,-P,0)$ the sum $\vec V_1+\vec V_2^{+-} = (2P,0,0)$ is associated with the tetragonal domain $4_xmm$ (Fig. \ref{fig1} (d)), the domain wall having the maximal isotropy subgroup $m_ym_z2_x$. For $\vec V_2^{--} = (-P,0,-P)$ the domain wall symmetry $m_xm_{yz}2_{\overline yz}$ is that of one of the twelve orthorhombic domain states $\vec V_1+\vec V_2^{--} = (0,P,-P)$ (Fig. \ref{fig1} (d)). For $\vec V_2^{++}=(P,0,P)$ the sum $\vec V_1+\vec V_2^{++} = (2P,P,P)$ yields a domain wall symmetry $m_{\overline yz}$ corresponding to the monoclinic domain state denoted IV in table \ref{tab1}, which is not stabilized in bulk BaTiO$_3$ although its presence has been disputed in literature \cite{Keeble2009}.
   
\begin{figure}[th]
\begin{center}
\includegraphics[width=0.48\textwidth]{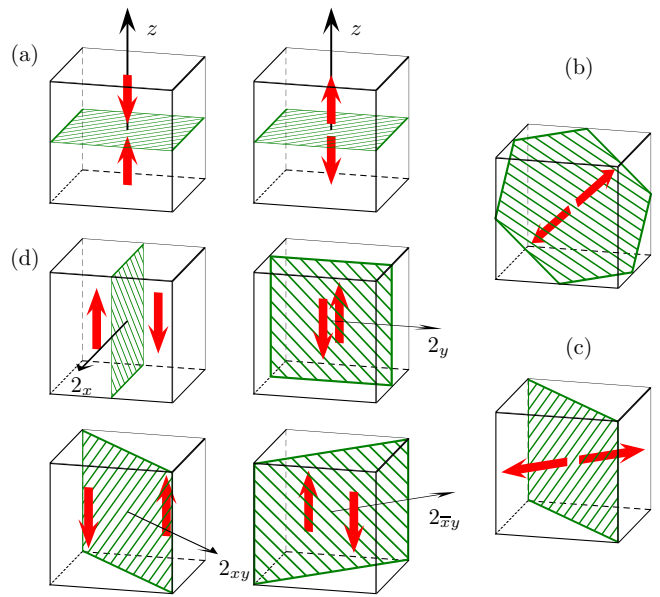}
\caption{Orientation of the polarization (red arrows) on each side of the 180$^\circ$ domain walls in the tetragonal ((a), (d)), rhombohedral (b) and orthorhombic (c) phases of BaTiO$_3$. The detailed meaning of the figures is in the text.}
\label{fig2}
\end{center}
\end{figure}

For determining the domain wall symmetries between the 180$^\circ$ ferroelectric domains of BaTiO$_3$, one has to take into account the reduction of the order-parameter space $E_3$ for a given orientation of the domains. Considering the tetragonal 180$^\circ$ domains represented by the $E_3$-vectors $\vec V_1 = (0,0,P)$ and $\vec V_2 = (0,0,-P)$ gives the sum $\vec V_1 + \vec V_2= (0,0,0)$ corresponding to the equilibrium value of the order-parameter in the \emph{parent ferroelastic domain} associated with $\vec V_1$ and $\vec V_2$, which has the tetragonal point-group symmetry $4_z/mmm$. This is the actual symmetry of the head-to-head or tail-to-tail 180$^\circ$ domain walls shown in Fig. \ref{fig2} (a). Following the same scheme the domain-wall separating 180$^\circ$ rhombohedral antiparallel domains such as $\pm(P,P,P)$ has the symmetry $R\overline 3^{xyz}m$ of the parent ferroelastic domain (Fig. \ref{fig2} (b), whereas the orthorhombic ferroelastic domain symmetry $mmm$ coincides with the symmetry of orthorhombic domain walls between $\pm(P,P,0)$ 180$^\circ$ domains (Fig. \ref{fig2} (c)). 

For tetragonal 180$^\circ$ domains separated by planes ($m_x$,$m_y$,$m_{xy}$,$m_{\overline xy}$) containing the four-fold rotation $4_z$, the symmetry of the domain walls is lowered when \emph{fixing} the orientation of one of the planes. This is consistent with our approach since lowering the $m\overline 3m$ symmetry to $4/mmm$ yields a decomposition of the order-parameter space $E_3=E_2+E_1$. In the two-dimensional order-parameter space $E_2$, spanned by the bases $\vec V_1=(\pm P,0),\vec V_2=(0,\pm P)$ or $\vec V_1=(\pm P,\pm P),\vec V_2=(\pm P,\mp P)$, the sum $\vec V_1 + \vec V_2$ provides the domain-wall orientations $\overline xy$, $xy$, $y$ and $x$ having the respective monoclinic symmetries $2_{xy}/m_{xy}$, $2_{\overline xy}/m_{\overline xy}$, $2_x/m_x$ and $2_y/m_y$. They correspond to the four 180$^\circ$ domain configurations shown in Fig. \ref{fig2} (d). In the one-dimensional order-parameter space $E_1$ the basic vector $\vec V$ coincides with the single domain state $\pm P$ of tetragonal symmetry $4_z/mmm$ which is the symmetry of the head-to-head and tail-to-tail domain walls (Fig. 2.(a)).

\section{Domain-wall symmetry in ferroelastic S\MakeLowercase{r}T\MakeLowercase{i}O$_3$ and L\MakeLowercase{a}A\MakeLowercase{l}O$_3$}

As a second example we describe the domain wall symmetries separating the ferroelastic domains in LaAlO$_3$ and SrTiO$_3$ which undergo \emph{improper} ferroelastic transitions \cite{Bueble1998,Scott1969} leading, respectively, to rhombohedral and tetragonal phases for different equilibrium values of the same three-dimensional order-parameter symmetry $(\eta_1,\eta_2,\eta_3)$ \cite{Stokes1988} (Table \ref{tab1}). In the cell-doubling transition $Pm\overline 3m(\vec a,\vec b,\vec c)\rightarrow R\overline 3c(\vec a+\vec b,\vec b+\vec c,\vec c+\vec a)$ of LaAlO$_3$, the order-parameter has an eight-fold degeneracy corresponding (1) to the equilibrium values I:$(\eta,\eta,\eta)$, II:$(-\eta,\eta,\eta)$, III:$(\eta,-\eta,\eta)$, IV:$(\eta,\eta,-\eta)$ associated with four ferroelastic domains $(e_{xy},e_{zx},e_{yz})$, $(-e_{xy},-e_{zx},e_{yz})$, $(-e_{xy},e_{zx},-e_{yz})$ and $(e_{xy},-e_{zx},-e_{yz})$, and (2) to the opposite values V:$(-\eta,-\eta,-\eta)$, VI:$(\eta,-\eta,-\eta)$, VII:$(-\eta,\eta,-\eta)$, VIII:$(-\eta,-\eta,\eta)$ representing \emph{antiphase domains} related to the loss of the cubic translations $\vec a$, $\vec b$ and $\vec c$ at the transition. The ferroelastic domain walls I--II, I--III, I--IV, II--III, II--IV and III--IV display two types of symmetries: (1) orthorhombic $m_xm_{yz}m_{\overline yz}$ for the domain wall I--II, associated with the sum $\vec V_\mathrm{I}+\vec V_\mathrm{II}=(0,2\eta,2\eta)$, coinciding with one of the domain states of the orthorhombic $Imma$ phase (Table \ref{tab1}), and (2) again orthorhombic $m_xm_ym_z$ for the domain wall II--III, given by $\vec V_\mathrm{II}+\vec V_\mathrm{III}=(0,0,2\eta)$, which is a maximal isotropy subgroup of the domain state $I4_z/mmm$ (Table \ref{tab1}). \emph{Antiphase domain walls}, i.e. boundaries between distinct antiphase domains within the same ferroelastic domain, correspond to the I--V, II--VI, III--VII and IV--VIII pairs displaying the $R\overline 3m$ symmetry of the rhombohedral ferroelastic domain.

A similar description can be given for the ferroelastic and antiphase domain walls in the tetragonal $I4/mcm(\vec a+\vec c,2\vec b,\vec c - \vec a)$ phase of SrTiO$_3$. Thus, the walls between the three ferroelastic domains have the orthorhombic symmetries $m_{xy}m_{\overline xy}m_z$, $m_{xz}m_{\overline xz}m_y$ and $m_{yz}m_{\overline yz}m_x$ corresponding to the point groups of the orthorhombic domain states $Imma$ (Table \ref{tab1}), whereas the symmetries of the three distinct antiphase domain walls have the ferroelastic domain symmetries $4_u/mmm$ ($u=x,y,z$). 

This result calls for comments in light of the recent experimental observation of piezoelectric resonance below 80~K in SrTiO$_3$ \cite{Salje2013}, interpreted as a signature of the polar character of the domain walls. In SrTiO$_3$ and LaAlO$_3$ the primary order-parameter symmetry associated to the tilt of octahedra always preserves the inversion center and therefore leads to domain walls with an intrinsic non-polar character. In contrast, the standard geometric approach \cite{Janovec2006} states that \emph{all} compatible ferroelastic walls are polar, irrespective of any ferroelectric instability. This apparent contradiction can be waived by recalling that this latter prediction is only related to the boundary conditions imposed on the internal domain-wall structure for a specific wall orientation, but is not related to the phase-transition mechanism. If this mechanism becomes more complex, with electric polarization also involved as an order-parameter as it is often assumed in studies on SrTiO$_3$, we expect that a possible polar character of the wall will show up in the intrinsic symmetry of the domain wall derived from the order-parameter description, thereby distinguishing SrTiO$_3$ from ordinary ferroelastics.

\section{Antiphase domain walls in G\MakeLowercase{d}$_2$(M\MakeLowercase{o}O$_4$)$_3$ and YM\MakeLowercase{n}O$_3$}

\begin{figure}[ht]
\begin{center}
\includegraphics[width=0.48\textwidth]{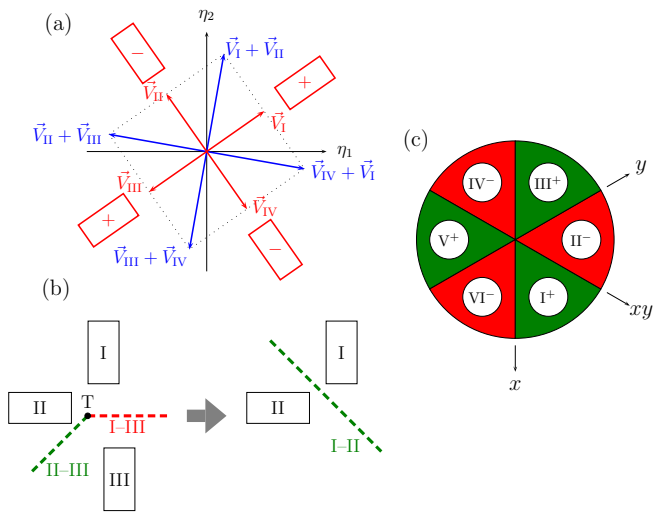}
\caption{(a) Orientation of the FF domains (red arrows and rectangles) and domain-wall vectors (blue arrows) for the four equilibrium domain states of GMO. The detailed meaning of the figure is in the text. (b) Figure illustrating the simultaneous creation of a FF domain wall (green hatched line I--II) and annihilation of an antiphase domain wall (red hatched line III--I) in GMO. (c) Cloverleaf domain pattern configuration observed in YMnO$_3$ \cite{Chae2010}. The six ferroelectric domains of alternating up-down polarization (green-red) emerge from a single six domain-state point, together with the annihilation of the domain walls between the three antiphase domains.}
\label{fig3}
\end{center}
\end{figure}

Although the domain-wall symmetry derived from the order-parameter symmetry often coincides with the symmetry resulting from the geometrical approach in terms of layer groups \cite{Janovec2006}, such coincidence is not always realized as, for example, at the improper ferroelectric-ferroelastic (FF) transition in gadolinium molybdate (GMO) \cite{Cross1968}. The two-component order-parameter $(\eta_1,\eta_2)$ describing the $P\overline 42_1m(\vec a,\vec b,\vec c)\rightarrow Pba2(\vec a-\vec b,\vec a+\vec b,\vec c)$ transition in GMO has a fourfold degeneracy corresponding to the equilibrium domain states I:$(\eta_1,\eta_2)$, II:$(-\eta_2,\eta_1)$, III:$(-\eta_1,-\eta_2)$ and IV:$(\eta_2,-\eta_1)$, with two opposite FF domains $\pm(P_z,e_{xy})$ (Fig. \ref{fig3} (a)). There are four types of FF domain walls (I--II, II--III, III--IV and I--IV), and two types of antiphase domain walls (I--III and II--IV). The four FF domains walls have the orthorhombic symmetry $mm2$, \emph{which is higher than the geometrically determined monoclinic symmetry $2$}. This is because the order-parameter space has the point-group symmetry $4$ containing only rotations about an axis perpendicular to the order-parameter plane $(\eta_1,\eta_2)$ \cite{Toledano1987}. Hence, this plane has no specific symmetry directions, the sums $\vec V_\mathrm{I}+\vec V_\mathrm{II}$, $\vec V_\mathrm{II}+\vec V_\mathrm{III}$, $\vec V_\mathrm{III}+\vec V_\mathrm{IV}$, $\vec V_\mathrm{IV}+\vec V_\mathrm{I}$ associated with the domain walls having the same orthorhombic symmetry as the domain states. The symmetry of the antiphase domain walls I--III and II--IV, given by $\vec V_\mathrm{I}+\vec V_\mathrm{III} = \vec V_\mathrm{II}+\vec V_\mathrm{IV} = (0,0)$, is the maximal isotropy subgroup $m_xm_y2_z$ of the parent symmetry $\overline 4m2$. 

The determination of the antiphase domain wall symmetries is derived from the primary order-parameter space following the same summing rule as for ferroic domain walls. The antiphase domain-walls in LaAlO$_3$, SrTiO$_3$ and GMO display the higher symmetry of the ferroelastic domains in which they form, due to the fact that they separate antiphase domains with opposite order-parameter components. However, this is not the most general case and antiphase domain-walls can have a lower symmetry than the domains they separate. Furthermore, the coexistence of ferroic and antiphase domain walls can lead to complex interactions that may result in the \emph{annihilation} or \emph{creation} of both types of domain walls for specific configurations of the domains \cite{Meleshina1974}. For example, in GMO the junctions between the antiphase domain wall I--III and the FF domain wall III--II can merge at a triple point (Fig. \ref{fig3} (b)) corresponding to the annihilation of the antiphase wall I--III and the creation of the FF domain wall I--II, i.e. I--III $+$ III--II $\rightarrow$ I--II. Reciprocally the junctions between the FF domain walls I--II and II--III can produce an annihilation of the FF domain walls and creation of an antiphase domain wall I--III: I--II $+$ II--III $\rightarrow$ I--III. A striking illustration of the annhilation-creation process is found in the vortex-like domain pattern of YMnO$_3$ \cite{Chae2010} which shows the existence of a six-domain-state point at which annihilation of the antiphase domains results in the creation of six adjacent ferroelectric domains (Fig. \ref{fig3} (c)). The two component order-parameter associated with the cell-tripled $P6_3/mmc(\vec a,\vec b,\vec c)\rightarrow P6_3cm(2\vec a+\vec b,\vec b-\vec a,\vec c)$ improper ferroelectric transition in YMnO$_3$ gives rise to a total of six domain states combining three antiphase domains, resulting from the loss of the paraelectric translations $(\vec a,\vec b)$, with two opposite ferroelectric 180$^\circ$ domains along $\vec c$. The corresponding cloverleaf domain pattern contains alternating $\pm$ and $\mp$ 180$^\circ$ ferroelectric domain walls merging at a single point \emph{but no antiphase domain walls}. 

\section{Summary and conclusion}

In summary, it has been shown on selected examples of transitions that the symmetries of ferroelectric, ferroelastic and antiphase domain walls can be directly derived from the symmetry of the corresponding adjacent domains in the order-parameter vector space. In some cases, such as the walls between the 180$^\circ$-domains in BaTiO$_3$, one has to take into account a reduction of the parent order-parameter space. In all cases the domain-wall symmetry is an isotropy subgroup coinciding with the sum of the vectors associated with adjacent domains. Although only the point-group symmetry of the domain-walls has been worked out, the procedure also provides the space-group symmetries of the domain walls from which one can deduce the Bravais lattice along the two crystallographic directions preserved by the orientation of the domain wall. In our illustrative examples the maximal symmetry induced by the primary order-parameter has been considered for the domain walls without taking into account specific constraints which may reduce their symmetry. Let us finally emphasize that our proposed theoretical approach of domain walls applies to higher-order ferroics in which higher rank macroscopic tensors emerge spontaneously \cite{Toledano1977} or to non-ferroic transitions \cite{Toledano1982} involving exclusively antiphase domains. By contrast the domain wall symmetry of magnetic ferroics and multiferroics will be described elsewhere as it requires taking into account time-reversal symmetry.

The authors are grateful to B. Mettout for very helpful discussions. This work was supported by the National Research Fund, Luxembourg (FNR/P12/4853155/Kreisel). P.T.
is grateful to the CRP Gabriel Lippmann for financial support during his stay as visiting scientist.


\end{document}